\documentclass[useAMS,usegraphicx]{mn2e}

\title[The AGN content of ULIRGs]{A quantitative determination of the AGN content in local ULIRGs through L-band spectroscopy
}
\author[G. Risaliti et al.]{G.~Risaliti,$^{1,2}$ M.~Imanishi,$^3$ and E.~Sani$^4$
\\
$^1$ INAF - Osservatorio Astrofisico di Arcetri, L.go E. Fermi 5, 50125 Firenze, Italy
{E-mail: risaliti@arcetri.astro.it}
\\
$^2$ Harvard-Smithsonian Center for Astrophysics, 60 Garden St. Cambridge, MA 02138 USA\\
$^3$ National Astronomical Observatory, 2-21-1, Osawa, Mitaka, Tokyo 181-8588, Japan\\
$^4$ Dipartimento di Astronomia, Universit\`a di Firenze, L.go E. Fermi 2, 50125 Firenze, Italy.
}

\begin{document}
\date{Released Xxxx Xxxxx XX}

\pagerange{\pageref{firstpage}--\pageref{lastpage}} \pubyear{2008}

\maketitle

\label{firstpage}

\begin{abstract}
We present a quantitative estimate of the relative AGN/starburst content in a sample
of 59 nearby ($z<0.15$)
infrared bright ULIRGs taken from the 1~Jy sample, based on infrared L-band (3-4~$\mu$m) spectra. 
By using diagnostic diagrams and
a simple deconvolution model, we show that at least 60\% of local ULIRGs contain an active nucleus,
but the AGN contribution to the bolometric luminosity is relevant only in $\sim15-20$\% of the sources.
Overall, ULIRGs appear to be powered by the starburst process, responsible for $>85$\% of the
observed infrared luminosity. The subsample of sources optically classified as LINERs (31 objects)
shows a similar AGN/starburst distribution as the whole sample, indicating a composite nature for
this class of objects.
We also show that a few ULIRGs, optically classified as starbursts,
have L-band spectral features suggesting the presence of a buried AGN. 
\end{abstract}

\begin{keywords}
galaxies: active - galaxies:starburst - infrared:galaxies
\end{keywords}

\section{Introduction}
 Ultraluminous Infrared
Galaxies (ULIRGs, L$_{IR}\sim L_{BOL} > 10^{12} L_\odot$) have been 
studied at all wavelengths, from radio to hard X-rays, with the primary
aim of  
estimating the relative contribution of accretion and star formation activity
to the bolometric luminosity.

Recently, L-band (3-4~$\mu$m) spectroscopy provided new powerful tools to
perform this study. 
Imanishi \& Dudley (2000), Risaliti et al.~(2006, hereafter R06), Imanishi et al.~(2006,
hearafter I06), Imanishi et al.~(2008),
Sani et al.~(2008) 
analyzed nearby bright ULIRGs from the IRAS Bright Galaxy Sample (Sanders et al. 2003)
and the IRAS 1~Jy sample (Kim \& Sanders~1998) and demonstrated that 
the AGN and starburst components can be disentangled in L-band spectra, 
using several indicators:\\
- 3.3~$\mu$m emission feature: this emission is due to Policyclic Aromatic Hydrocarbon (PAH) molecules,
and is prominent in starburst-dominated sources (equivalent width EW$\sim110$~nm), while it is
weak or absent in AGN-dominated sources. 
This observational result is motivated by the different intrinsic emission
of AGNs and starbursts: the intense X-ray radiation of the former is believed to destroy the PAH
molecules, which can instead survive in the starburst radiation field.
\\
- 3.4~$\mu$m absorption feature: in ULIRGs, the presence of this feature with an optical depth $\tau_{3.4}>0.2$
is observationally associated to heavily obscured AGNs. 
This is due to the centrally concentrated emission of AGNs, which can be effectively
covered by large columns of dust. Imanishi \& Maloney (2003) demonstrated that a large-scale source, such
as a starburst, interspersed within the dust responsible for absorption,  cannot have a 3.4~$\mu$m absorption feature
with $\tau>0.2$.\\
- Continuum slope: ULIRGs containing heavily obscured AGNs show a much redder L-band continuum
than unobscured AGNs and starbursts. This is due to the reddening effect of the covering dust. If the continuum
is modeled with a simple power law in the $\lambda$-$f_\lambda$ plane, i.e. $f_\lambda$$\propto$$\lambda^{\Gamma}$,
the typical values for the slope $\Gamma$ are $\Gamma$$\sim$-0.5 for unobscured AGNs and pure starburst,
while $\Gamma$$>$1 for heavily obscured AGNs (R06).
\\
- Bolometric ratios: the intrinsic ratio between the $3~\mu$m and bolometric luminosity
is $\sim100$~times higher in pure AGNs than in starbursts (R06). The measured ratio in ULIRGs is therefore
in itself an indicator of the possible presence of an AGN component.
\\  
The power of L-band spectroscopic analysis lays in the capability of detecting heavily obscured AGNs,
which are missed at other wavelenghs. The combination of the properties listed above implies that 
an AGN with an L-band optical depth $A_L\sim1-5$
(corresponding to $A_V$$\sim$20-100, i.e. completely absorbed from the UV to the near-IR) can still
be detectable in the L-band spectrum of ULIRGs, even when its contribution
to the bolometric luminosity is negligible. \\
A quantitative application of the above diagnostics has been proposed by R06,
who assumed that the L-band spectra can be reproduced as the combination of a fixed starburst template 
and an AGN template with variable dust absorption. As a consequence, the model has two free 
parameters: the fraction of the AGN contribution to the L-band luminosity, $\alpha$, and
the optical depth $\tau_L$ of the dust absorbing the AGN component. These two parameters can
be estimated from the observed continuum slope, $\Gamma$, and the equivalent width $EW_{3.3}$ of
the 3.3~$\mu$m PAH emission feature. 

In this paper we present the application of this method to a complete sample of 59 ULIRG
with available L-band spectroscopic observations. The sample consists of all the ULIRGs
in the IRAS 1~Jy sample with redshift $z<0.15$. The optical classification (Veilleux et al.~1999)
is AGN for 14 sources (25\%), starburst for 12 sources (22\%) and LINER for 31 sources (56\%).
Two sources are not optically classified.
Our work is mostly based on published data, which already provided an AGN/starburst classification,
based on the diagnostic features summarized above. The significant improvement presented here
consists in moving from a simple ULIRGs classification as starburst or AGN, to a quantitative estimate of 
the contribution of the two components to the bolometric luminosity.

A similar approach has also been successfully used by Nardini et al.~(2008) on 
{\em Spitzer}/IRS 5-8~$\mu$m spectra of a sample of local ULIRGs largely overlapping
with the one presented here. 
We will discuss our results in comparison with those of Nardini et al.~2008 in Section~3.


\section{Data Analysis}

The observations in the northern hemisphere have been performed with the SUBARU (45 sources)
and IRTF (6 sources) telescopes, and have been presented in I06. In the southern
hemisphere the observations were made with the VLT (eight sources). Four of these sources have been
presented by R06 and Risaliti et al. 2006B, while the remaining four were observed in early 2006
with ISAAC, with the same configuration as for the first eight sources, and reduced
following the steps described in R06. We refer to R06 and I06 for
a detailed description of the data reduction and a visual analysis of the spectra.
 
We performed a homogeneous analysis of all the sources in our sample, fitting each $f_\lambda$ spectrum with 
a standard $\chi^2$ minimization using stated observational uncertainties. The adopted model consists of a  
power law continuum, and a Gaussian emission feature at rest frame wavelength of 3.3~$\mu$m.
For each source, the spectral interval used for the fit is $\lambda=3.25\times(1+z)-4.1~\mu$m, i.e. the
region containing the 3.3~$\mu$m emission feature and the continuum at longer wavelengths up to the
end of the L-band. We excluded the wavelengths shortward of the PAH emission feature to avoid
regions of bad atmospheric transmission, and possible distortions of the continuum slope due to
ice absorption features (see R06 and I06 for more details).
In a few spectra an hydrocarbon absorption feature at rest frame wavelength $\lambda_{rest}=3.4-3.5~\mu$m is clearly present.
In these cases we added a Gaussian absorption component in our fits. In all the other cases we checked that
this extra component is not required statistically.
In Fig.~1 we show a few examples of the data and spectral fitting for representative
objects in our sample.
\begin{figure}
\includegraphics[width=8.5cm]{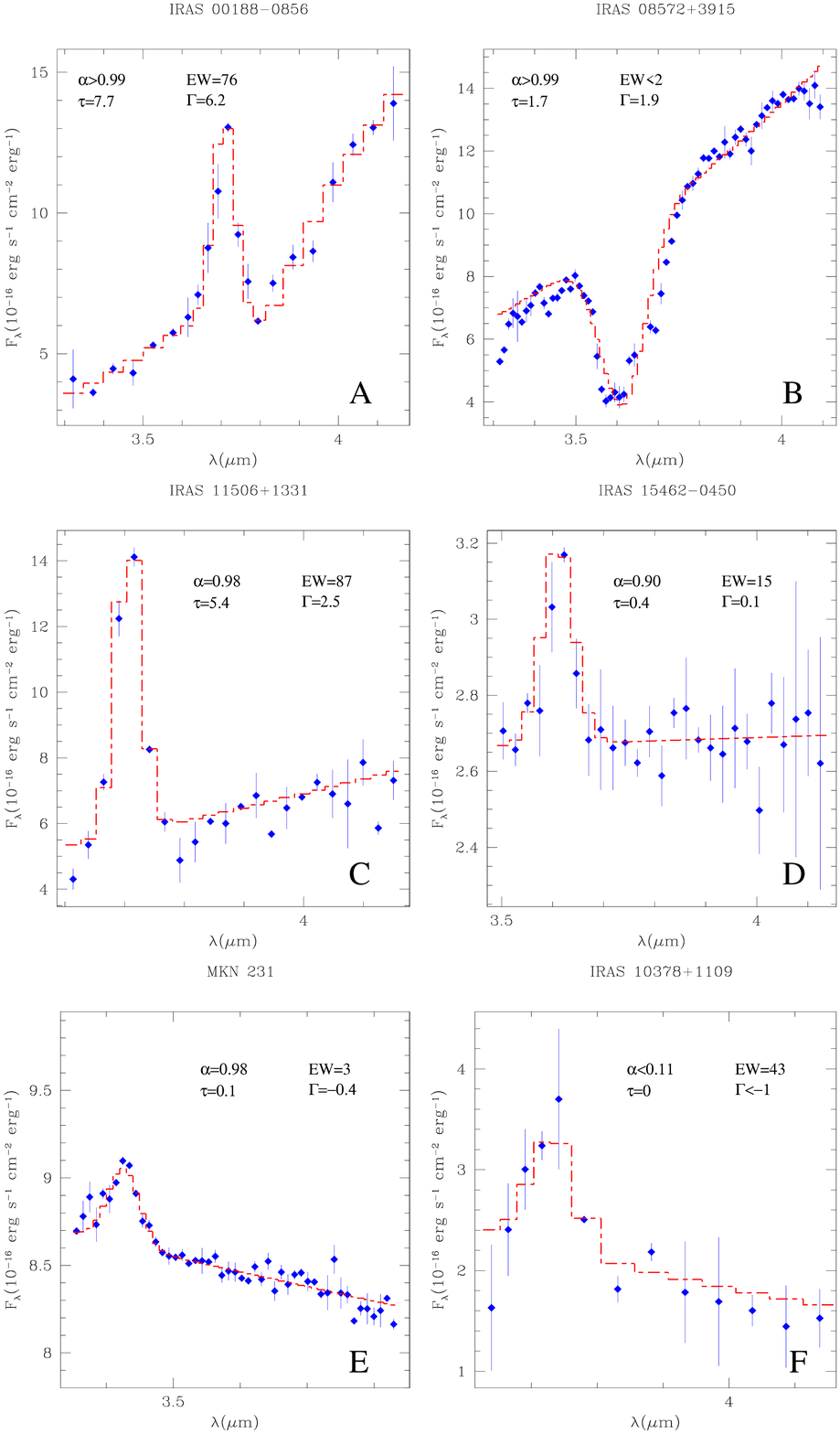}
\caption{
Six examples of our data and best fit models, including objects among the highest ({\bf B}) and lowest ({\bf F})  
S/N.
{\bf A}, {\bf B} and {\bf C}: sources {\em intrinsically} dominated by a heavily obscured AGN. The starburst component is
however strong in the {\em observed} spectra in {\bf A} and {\bf C}. Note that {\bf C} is optically classified as starburst.
{\bf D} and {\bf E}: objects containing a weakely obscured AGN, which is dominant in the L-band, but not in
the Bolometric luminosity (See Table~1). {\bf F}: Starburst-dominated object.
} 
\end{figure}
\begin{table*}
\begin{scriptsize}
\centerline{\begin{tabular}{lcccccccc}
Source (instrument$^a$)& 
$\Gamma$               &    
EW                     &
$\tau$                &
$\alpha_L$             & 
$\alpha_{BOL}$         & 
$\tau_{3.4}$           &
Cl(L)$^b$              &
Cl(Sp)$^c$            \\
\hline
 IRAS 00188-0856 (S)&  6.2$\pm$0.1&   76$\pm$   13&     7.7$\pm$   1.4& 0.995$\pm$ 0.003& 0.91$\pm$ 0.06 &0.36$\pm$0.06 & AGN* & AGN \\
 IRAS 02411+0353E (V)&-2.4$\pm$2.2&  177$\pm$   51&      0$^e$        & $<0.1^e$      & $<10^{-3e}$    & 0$^f$          & --   & SB  \\
 IRAS 02411+0353W (V)&-2.5$\pm$3.0&  183$\pm$   80&      $<3$         & $<0.96$      & $<0.18$       & 0$^f$            & --   & SB  \\
 IRAS 03250+1606 (S)& -1.8$\pm$1.1&   83$\pm$   22&      0$^d$        & $<$0.52       & $<$0.01        & 0$^f$          & AGN* & SB  \\
 IRAS 04103-2838 (V)& -0.1$\pm$0.2&   59$\pm$    6&     0.7$\pm$   0.4& 0.63$\pm$0.07 & 0.02$\pm$0.005 & 0$^f$          & --   & AGN*\\
 IRAS 08572+3915 (I)&  1.9$\pm$0.1&    1$\pm$    1&     1.7$\pm$   0.1& $>0.99$       & $>0.71$        &0.88$\pm$0.05   & AGN  & AGN \\
 IRAS 09039+0503 (S)&  0.3$\pm$0.8&  257$\pm$   88&      0$^e$        & $<0.1^e$       & $<10^{-3e}$    & 0$^f$         & AGN* & AGN*\\   
 IRAS 09116+0334 (S)& -1.9$\pm$0.3&   77$\pm$   18&      0$^d$        & 0.3$\pm0.2$   & 0.02$^{+0.015}_{-0.015}$& 0$^f$ & AGN* & SB  \\
 IRAS 09539+0857 (S)& -0.5$\pm$0.2&  101$\pm$   21&     $<$0.2        & $<$0.09       & $<$0.01        & 0$^f$          & SB   & AGN*\\
 IRAS 10378+1108 (S)& -2.8$\pm$1.2&   43$\pm$   24&      0$^d$        & 0.6$\pm0.2$   & 0.02$_{-0.01}^{+0.02}$ & 0$^f$  & SB   & AGN*\\
 IRAS 10485-1447 (S)&  3.4$\pm$0.2&   41$\pm$    3&     3.6$\pm$   0.3& 0.98$\pm$ 0.01& 0.37$\pm$ 0.03 & 0.25$\pm$0.08  & AGN* & AGN*\\
 IRAS 10494+4424 (S)& -0.5$\pm$0.2&  126$\pm$   15&     $<$0.2        & $<$0.11       & $<$0.01        & 0$^f$          & AGN* & SB  \\
 IRAS 11095-0238 (S)&  1.5$\pm$0.1&  127$\pm$   26&     5.7$\pm$   3.9& $>$0.95       & $>0.20$        & 0$^f$          & SB   & AGN \\
IRAS 12112+0305NE (V)& 0.1$\pm$1.0&   69$\pm$   30&     $<3$          & 0.63$\pm$ 0.35& $<0.04$        & 0$^f$          & SB   & SB  \\
IRAS 12112+0305SW (V)&-0.9$\pm$0.7&   $<300$      &      --           &    --         &   --           & 0$^f$          & --   & --  \\
 IRAS 12359-0725 (S)&  2.3$\pm$0.3&   83$\pm$   17&     4.6$\pm$   1.8& 0.97$\pm$ 0.02& 0.25$\pm$ 0.15 & 0$^f$          & AGN* & AGN*\\
 IRAS 12127-1412 (S)&  2.1$\pm$0.1&   21$\pm$    6&     2.0$\pm$   0.1& 0.97$\pm$ 0.01& 0.25$\pm$ 0.06 & 0$^f$          & AGN  & AGN \\
 IRAS 13335-2612 (V)&  2.0$\pm$1.5&  220$\pm$  146&      0$^e$        & $<0.1^e$     & $<10^{-3e}$    & 0$^f$           & --   & SB  \\
IRAS 14252-1550E (S)&  0.1$\pm$1.6&   85$\pm$   32&      --           &    --         &   --           & 0$^f$          & --   & --  \\
IRAS 14252-1550W (S)&  0.1$\pm$0.3&   91$\pm$   20&     $<$1.8        & 0.55$\pm$ 0.30& 0.01$\pm$ 0.008& 0$^f$          & AGN* & SB  \\
 IRAS 14348-1447 (V)&  0.4$\pm$0.4&  110$\pm$   30&     1.2$\pm$   1.1& $<$0.72       & $<$0.03        & 0$^f$          & AGN* & AGN*\\
  Arp        220 (I)&  0.4$\pm$0.2&   79$\pm$    9&     1.8$\pm$   0.7& 0.69$\pm$ 0.10& 0.02$\pm$ 0.01 & 0$^f$          & SB   & AGN*\\
 IRAS 16090-0139 (S)&  2.5$\pm$5.8&   $<180$      &      --           &    --         &   --           & 0$^f$          & AGN* & AGN*\\
 IRAS 16468+5200 (S)& -2.8$\pm$1.6&  184$\pm$   70&      --           & $<0.54$       & $<0.01$        & 0$^f$          & SB   & AGN*\\
 IRAS 16487+5447 (S)& -2.0$\pm$1.4&  126$\pm$   37&      --           &    --         &   --           & 0$^f$          & AGN* & SB  \\
 IRAS 17028+5817 (S)& -2.1$\pm$0.2&   83$\pm$   11&      0$^d$        & 0.24$\pm$0.10 & 0.003$\pm$0.001& 0$^f$          &AGN*  & SB  \\
 IRAS 17044+6720 (S)&  1.3$\pm$0.1&   18$\pm$    4&     1.4$\pm$   0.1& 0.95$\pm$ 0.01& 0.17$\pm$ 0.03 & 0.14$\pm$0.09  & AGN  & AGN \\
 IRAS 21219-1757 (S)& -0.7$\pm$0.1&    1$\pm$  0.5&      0$^d$        & 0.99$\pm$ 0.005& 0.47$\pm$ 0.10& 0$^f$          & --   & --  \\
 IRAS 21329-2346 (S)& -0.4$\pm$0.7&   79$\pm$   16&     $<1.5$        & $<0.74$       & $<0.02$        & 0$^f$          & AGN* & AGN*\\
 IRAS 23234+0946 (S)&  0.6$\pm$0.1&  146$\pm$   14&      0$^e$        & $<0.1^e$     & $<10^{-3e}$    & 0$^f$           & SB   & SB  \\
 IRAS 23327+2913 (S)&  0.1$\pm$0.8&   52$\pm$   16&     $<2.0$        & 0.70$\pm$ 0.20& 0.023$\pm$0.017& 0$^f$          & SB   & AGN*\\
\hline
 IRAS 10190+1322E(S)& -0.5$\pm$0.1&   82$\pm$    3&     $<$0.2        & 0.26$\pm$ 0.06& 0.003$\pm$0.001& 0$^f$          & SB   & SB  \\
 IRAS 10190+1322W(S)& -2.2$\pm$0.4&   79$\pm$   20&      0$^d$        & $<0.46$       & $<0.02$        & 0$^f$          & SB   & SB  \\
 IRAS 11387+4116 (S)& -1.1$\pm$0.4&  106$\pm$   31&      0$^d$        & $<$0.45       & $<$0.02        & 0$^f$          & SB   & SB  \\
 IRAS 11506+1331 (S)&  2.5$\pm$0.2&   87$\pm$   10&     5.4$\pm$   1.3& 0.98$\pm$ 0.01& 0.35$\pm$ 0.12 & 0$^f$          & AGN  & AGN \\
 IRAS 13509+0442 (S)&  0.2$\pm$1.3&  157$\pm$   35&     $<3.4$        & $<$0.97       & $<$0.28        & 0$^f$          & SB   & SB  \\
 IRAS 13539+2920 (S)& -0.6$\pm$0.6&   94$\pm$   20&     $<2.8$        & $<$0.73       & $<$0.03        & 0$^f$          & SB   & SB  \\
 IRAS 14060+2919 (S)& -0.2$\pm$0.2&  185$\pm$   15&      0$^e$        & $<0.1^e$     & $<10^{-3e}$    & 0$^f$           & SB   & SB  \\
 IRAS 15206+3342 (S)&  0.2$\pm$0.2&   71$\pm$   14&     1.1$\pm$   0.8& 0.64$\pm$ 0.15& 0.02$\pm$ 0.01 & 0$^f$          & SB   & SB  \\
 IRAS 15225+2350 (S)&  1.3$\pm$0.6&   28$\pm$   13&     1.5$\pm$   0.8& 0.93$\pm$ 0.04& 0.12$\pm$ 0.07 & 0$^f$          & AGN* & AGN*\\
 IRAS 16474+3430 (S)&  0.5$\pm$0.4&  103$\pm$   29&     4.8$\pm$  13.4& 0.90$\pm$ 0.48& 0.08$\pm$ 0.36 & 0$^f$          & AGN* & SB  \\
 IRAS 20414-1651 (S)& -1.4$\pm$1.0&   77$\pm$   18&     1.5$\pm$   3.4& $<$0.66       & $<$0.02        & 0$^f$          & SB   & SB  \\
 IRAS 21208-0519 (S)& -0.4$\pm$3.1&  169$\pm$  102&      --           & --            &  --            & 0$^f$          & SB   & SB  \\
\hline
 IRAS 05189-2524 (V)&  0.1$\pm$0.1&    7$\pm$    1&     0.4$\pm$   0.1& 0.96$\pm$ 0.01& 0.18$\pm$ 0.01 & 0$^f$          & AGN  & AGN \\
 IRAS 08559+1053 (S)& -0.2$\pm$0.1&   13$\pm$    4&     0.3$\pm$   0.1& 0.91$\pm$ 0.03& 0.09$\pm$ 0.03 & 0$^f$          & AGN  & --  \\
 IRAS 12072-0444 (S)&  1.2$\pm$0.2&   86$\pm$   19&     3.5$\pm$   1.9& 0.90$\pm$ 0.08& 0.08$\pm$ 0.06 & 0.57$\pm$0.11  & AGN  & --  \\
  Mrk        273 (I)&  0.1$\pm$0.1&   31$\pm$    5&     0.5$\pm$   0.2& 0.81$\pm$ 0.04& 0.04$\pm$ 0.01 & 0$^f$          & AGN  & AGN \\
 IRAS 13443+0802 (S)& -0.6$\pm$1.1&  111$\pm$   56&     $<2.0$        & $<0.5$        &$<0.01$         & 0$^f$          & SB   & --  \\
  PKS    1345+12 (I)&  0.7$\pm$0.1&    7$\pm$    3&     0.9$\pm$   0.1& 0.97$\pm$ 0.01& 0.25$\pm$ 0.08 &0$^f$           & AGN  & AGN \\
 IRAS 15130-1958 (S)&  0.2$\pm$0.1&    8$\pm$    1&     0.5$\pm$   0.1& 0.95$\pm$ 0.01& 0.17$\pm$ 0.05 &0$^f$           & AGN  & --  \\
 IRAS 17179+5444 (S)&  1.3$\pm$0.3&   24$\pm$   13&     1.5$\pm$   0.4& 0.94$\pm$ 0.04& 0.14$\pm$ 0.09 &1$\pm$0.5       & AGN  & --  \\
  Mrk       1014 (I)& -0.7$\pm$0.1&    8$\pm$    1&      0$^d$        & 0.91$\pm$ 0.01& 0.09$\pm$ 0.01 & 0$^f$          & AGN  & --  \\
 IRAS 07599+6508 (S)& -0.3$\pm$0.1&    1$\pm$    1&     0.2$\pm$  0.05& $>0.99$       & $>0.7$         & 0$^f$          & AGN  & --  \\
 IRAS 11598-0112 (S)& -0.3$\pm$0.2&    5$\pm$    3&     0.1$\pm$   0.1& 0.96$\pm$ 0.03& 0.19$\pm$ 0.10 & 0$^f$          & AGN  & --  \\
  Mrk        231 (I)& -0.4$\pm$0.1&    3$\pm$    0.5&   0.1$\pm$  0.05& 0.98$\pm$ 0.01& 0.28$\pm$ 0.02 & 0$^f$          & AGN  & AGN \\
 IRAS 15462-0450 (S)&  0.1$\pm$0.2&   15$\pm$    5&     0.4$\pm$   0.2& 0.90$\pm$ 0.04& 0.09$\pm$ 0.03 & 0$^f$          & AGN  & --  \\
 IRAS 12127-1412 (S)&  2.0$\pm$0.1&   21$\pm$    6&     2.0$\pm$   0.1& 0.97$\pm$ 0.01& 0.25$\pm$ 0.06 & 0.25$\pm0.07$  & AGN  & AGN \\
\hline
 IRAS 14197+0813 (S)& -0.6$\pm$0.5&  130$\pm$   16&     $<$0.9        & $<$0.67       & $<$0.02        & 0$^f$          & --   & --  \\
 IRAS 14485-2434 (S)&  0.7$\pm$0.1&   65$\pm$    8&     1.7$\pm$   0.3& 0.80$\pm$ 0.04& 0.04$\pm$ 0.01 & 0$^f$          & AGN  & --  \\
 \hline 
\end{tabular}}
\caption{
Results from AGN-SB decomposition as described in the text. The horizontal lines divide
the sources in groups according to their optical classification: LINERs (first group), starbursts (second group),
AGN (third group) and unclassified (the last two sources). 
The error on $\alpha_{BOL}$ does not take into account the
uncertainties on the L-band to bolometric ratios. A more realistic estimate can be estimate from
the scatter in Fig.~5B (see text for more details).
$^a$: Telescope/instrument 
of the observation: S: Subaru; V: VLT-ISAAC; I: IRTF. $^b$: Classification from ``standard'' L-band diagnostics,
(from I06), based on the PAH EW and the presence of absorption features. An asterisk indicates
a weak evidence for the presence of the AGN. $^c$: Classification
based on {\em Spitzer}-IRS spectroscopy (from Imanishi et al.~2007). $^d$: Absorption is assumed to
be $\tau=0$, due to the steep continuum ($\Gamma<-0.5$ at $>$90\% confidence level). $^e$: Values and upper limits
obtained assuming the source is pure starburst, due to the high equivalent width of 
the 3.3$\mu$m feature (EW$>$110~nm
at $>90$\% confidence level). $^f$: 3.4~$\mu$m absorption not required to adequately fit the spectrum.
}
\end{scriptsize}
\end{table*}
In Table~1 we list the best fit parameters $\Gamma$ and $EW_{3.3}$ for each source. A third
important parameter which can be directly inferred from the analysis is the ratio between the 
luminosity in the L-band and the bolometric one, obtained from IRAS fluxes.
As shown in R06, these three parameters are enough to provide a classification of the sources with 
respect to their AGN/starburst content. This is illustrated in Fig.~2 and~3. 
In Fig.~2 we show the position of our sources in the $\Gamma$-$EW_{3.3}$ plane. We also
show the regions occupied by obscured AGN, unobscured AGN, and starbursts, based on the 
calibration done in R06. 
Fig.~3 shows the measured L-band to bolometric flux ratio, $R_L$=$\lambda f_\lambda$(3.3~$\mu$m)/F(8-1000~$\mu$m) 
versus $EW_{3.3}$. We note that 
$R_L$ is in principle a rather powerful indicator,
given the huge difference between
the L-band to bolometric ratios for pure AGNs and pure starbursts (of a factor of about 100, R06).
However
it is not capable of giving an absolute estimate of the relative contributions of
the AGN/SB components, because of the degeneracy introduced by the extinction of the
AGN component: the same ratio can be obtained with a faint, unobscured AGN mixed with a powerful
starburst, or with
a much more luminous, even bolometrically dominant, but heavily obscured AGN. 
However, on average we observe an anticorrelation between this AGN indicator and
the 3.3~$\mu$m PAH starburst indicator (Fig.~3).

The second part of our analysis consists of a quantitative estimate of
the physically relevant parameters for our sources, i.e. the absorption of the AGN component, $\tau$,
and the continuum fraction of emission due to the AGN, $\alpha$ (estimated at 3.3~$\mu$m). We use the simple
method described in R06: the total spectrum is reproduced as a combination of an AGN power law component, 
$f_{AGN}=C_1 \lambda^\Gamma_{AGN}$, absorbed by dust following the extinction law $\tau\propto\lambda^{-1.75}$
(Cardelli et al.~1989), and a starburst component, consisting of a power law continuum 
$f_{SB}=C_2 \lambda^\Gamma_{SB}$ and a Gaussian emission line at $\lambda=3.3 \mu$m with a fixed
equivalent width $EW_{SB}$ with respect to the starburst continuum component.
Following the results obtained for bright ULIRGs in R06, we chose the values $\Gamma_{AGN}=-0.5$, $\Gamma_{SB}=-0.5$,
and $EW_{SB}=110$~nm for the AGN and SB templates.

Using this model, with simple algebra we can obtain the observed parameters $\Gamma$ and $EW_{3.3}$
as a function of 
the extinction $\tau$ and the AGN fraction $\alpha=C_1/(C_1+C_2)$.
$\tau$ and $\alpha$ can then be obtained simply inverting these equations, as shown in R06. The final
result is:
\begin{equation}
\tau_L=\frac{EW_{SB}\Gamma-EW_{3.3}}{\beta(EW_{SB}-EW_{3.3})}
\end{equation}
\begin{equation}
\alpha=\frac{EW_{SB}-EW_{3.3}}{(EW_{SB}-EW_{3.3})+EW_{3.3}*e^{-\tau_L}}
\end{equation}
where $\beta$ is the index of the extinction power law ($\beta$=1.75). 
The errors on $\alpha$ and $\tau$ can be directly estimated from the above equations,
as shown in the Appendix of R06.

In five cases the errors on the observed parameters are too high to obtain significant
estimates of $\tau$ and $\alpha$. For all the other sources the results are listed in Table~1.

The application of this model to the measured parameters listed in Table~1 is not straightforward for all sources:
in several cases the best fit values and the 90\% confidence intervals of one or both of $EW_{3.3}$ and $\Gamma$ 
are outside the ``allowed range'' of our
model, $\Gamma<\Gamma_{min}=$-0.5 and $EW_{3.3}>EW_{max}$=110~nm. 
In these cases our approach is the following:\\
$\bullet$ If $\Gamma<-0.5$ at  $>90$\% statistical significance, we assume $\tau=0$ and we estimate
the relative AGN fraction using $EW_{3.3}$: $\alpha=1-EW_{3.3}/(110~{\rm nm}$); \\
$\bullet$ If $EW_{3.3}>110$~nm and $\Gamma$ is not higher than the pure starburst value $\Gamma_{SB}=\Gamma_{min}=-0.5$
we assume the source is predominantly powered by a pure starburst and we put an upper limit of 1\% to the AGN contribution. 
We note that in no case $EW_{3.3}>110$~nm and $\Gamma>-0.5$ at a 90\% confidence level.
The values of $\alpha$ and $\tau$ so obtained are shown in Table~1. 

\begin{figure}
\includegraphics[width=8.5cm]{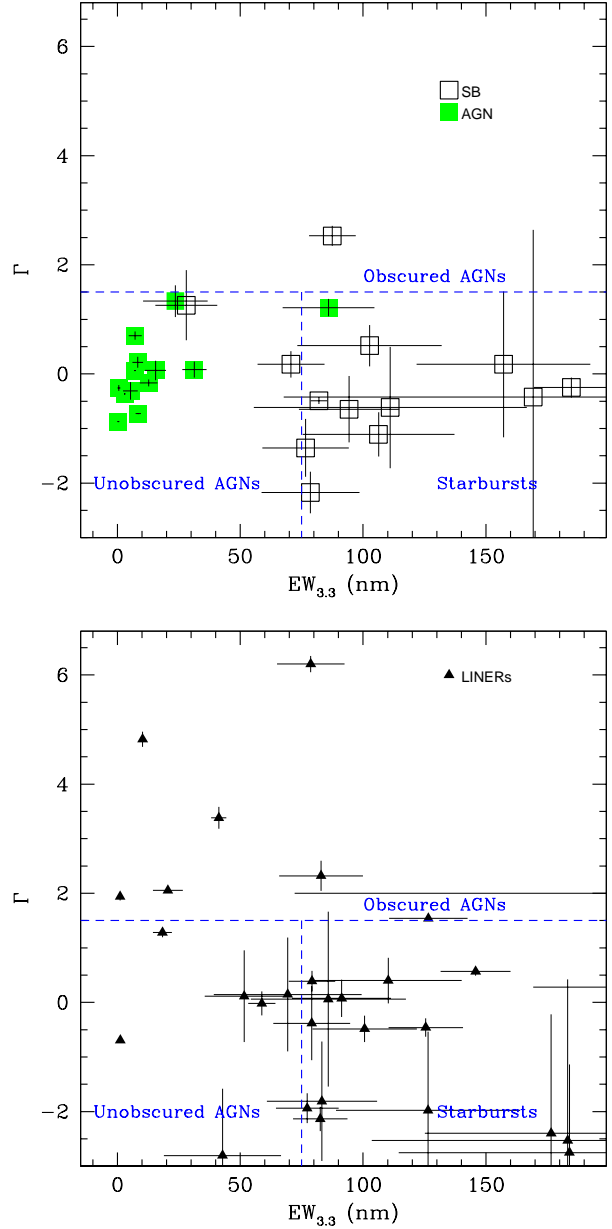}
\caption{3-4~$\mu$m continuum slope (in a $\lambda-f_\lambda$ spectrum) versus Equivalent Width of the 3.3~$\mu$m PAH
emission feature for the sample of $z<0.15$ ULIRGs in the 1~Jy sample (Veilleux et al.~1999). The dashed lines
mark the starburst, unobscured AGN and AGN zones according to the results of Risaliti et al.~2006. The sample
is split into two different panels for clarity, depending of the optical classification of the objects, marked as different symbols.
}
\end{figure}
\begin{figure}
\includegraphics[width=8.5cm]{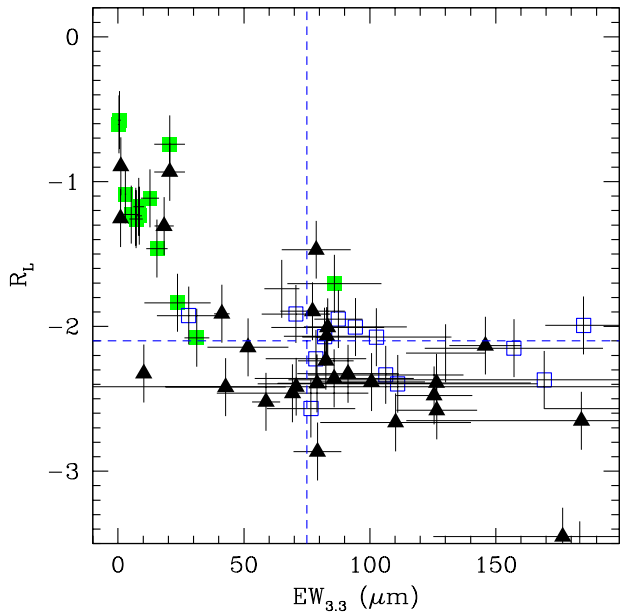}
\caption{ Ratio between L-band and bolometric luminosity, $R_L$, 
versus equivalent width of the 3.3~$\mu$m PAH emission feature. We adopt 20\% errors on $R_L$,
estimated from the average IRAS and L-band absolute calibration uncertainties.
The dashed lines mark the starburst and AGN zones
according to R06. Symbols are the same as in Fig.~2. 
}
\end{figure}

The final step in our analysis is the
estimate of the fraction of the bolometric emission due to the AGN, $\alpha_{BOL}$.
This quantity is related to the AGN fraction in the L-band, $\alpha$, through  two new
parameters, i.e. the ratio between the L-band and bolometric luminosities for pure AGN and pure SB, $R_{AGN}$
and $R_{SB}$. Simple calculations show that:
$\alpha_{bol}=\alpha/(\alpha+K(1-\alpha))$
where $K=R_{AGN}/R_{SB}$.
Following the results obtained for bright sources (R06) we use $R_{AGN}=0.2$ and $R_{SB}=0.002$.
The factor $K=100$ in the above equation imply that the bolometric contribution of the AGN can be small even
in those cases where the AGN component is dominant in the L-band. The values of $\alpha_{bol}$ estimated using
the above equation are listed in Column~6 of Table~1. The uncertainties related to the factor $K$ (log($K$)$\sim2$$\pm$0.5, R06) are discussed in Section~3.3.

\section{Discussion}
The application of our model to the 59 $z<0.15$ ULIRGs in the 1~Jy sample of Veilleux et al.~(1999)
provided 
%
a quantitative estimate of the AGN/starburst contribution in ULIRGs, and
in particular in those optically classified as LINERs.

Here we (1) review and discuss our results, (2) compare them
with the optical spectroscopic classification, and
with the other multi-wavelength AGN/starburst indicators summarized in the Introduction, and (3) discuss
the uncertainties of our approach.
\subsection{The AGN content in ULIRGs}
The most relevant result of the present work is a quantitative estimate of the
contributions of the two energy sources to the luminosity of ULIRGs.
The estimated AGN fractions in the L-band and in the total emission are summarized in the
histograms in Fig.~4. The comparison between the two distributions shows the power
of our diagnostics: the AGN, when present, dominates the L-band emission in
most cases, even in objects where its contribution to the total luminosity is negligible.
We note that our estimates strongly depend on the modelization of the starburst/AGN emission. In 
particular, we assume a fixed starburst template, with no free parameter. We expect this to be
a simplification of a more complex spectral behaviour, in at least two respects: a) the
intrinsic equivalent width of the 3.3~$\mu$m emission feature (assumed to be EW$_{3.3}=110$~nm in
our model) shows a significant spread
among pure starbursts, and b) the observed starburst emission could be affected by some 
reddening/absorption, as suggested by the large spread of the ratio between
flux of the 3.3~$\mu$m feature and the bolometric emission in pure starbursts (I06).
Therefore, we believe our analysis provides correct results in a statistical sense for the whole sample,
but can be affected by larger uncertainties than estimated in Table~1. We will discuss this issue
in more detail in Section~3.3.\\
Overall, our analysis shows that:\\
- An AGN is present in most ($>60$\%) ULIRGs. This fraction drops to $>30$\% for LINERs
(however, in several objects with low S/N the upper limit for the AGN fraction is quite high,
see Table~1). \\
- The contribution of AGNs to the L-band emission is $>50$\%, while the fraction
of the bolometric luminosity of ULIRGs due to AGNs is $<20$\%. Restricting to LINERs,
the fraction of AGN contribution is $<30$\% in the L-band, and $<15$\% bolometrically.\\
- In a few LINERs the presence of an heavily obscured AGN is bolometrically significant. On the other
side there is an indication that most of the heavily obscured AGN are LINERs (this
is based only on $\sim5-8$ objects, so it needs to be confirmed with larger samples).
This finding confirms the
heterogeneous nature of this class of objects, which are on average dominated by starburst emission,
but can host powerful buried AGNs.\\
- In our model, QSO class luminous buried AGNs are found
   only in a small fraction of ULIRGs classified optically
   as LINERs or starbursts.
\subsection{Comparison with other AGN indicators}
An AGN/SB classification for most of our sources has been presented, based on observations
at other wavelength, or on a different analysis of L-band data. \\
{\bf Optical:}
Fig.~2 shows the classification of our sources in a continuum slope-PAH EW plot, for 
each optical class: AGN, starburst, and LINER. AGNs and starbursts are clearly separated,
and are located in the regions found by R06. Two exceptions are found among optically
classified starbursts: two sources, IRAS~20414-1651 and IRAS 11506+1331 (Panel~C in Fig.~1) 
show 
signatures of the presence of absorbed AGNs, which were missed by studies at other
wavelengths. LINERs are spread throughout the plot, confirming their composite nature,
as quantitatively shown in Table~1 and Fig.~4.\\
{\bf L-band absorption features:} 
\begin{figure}
\includegraphics[width=8.5cm]{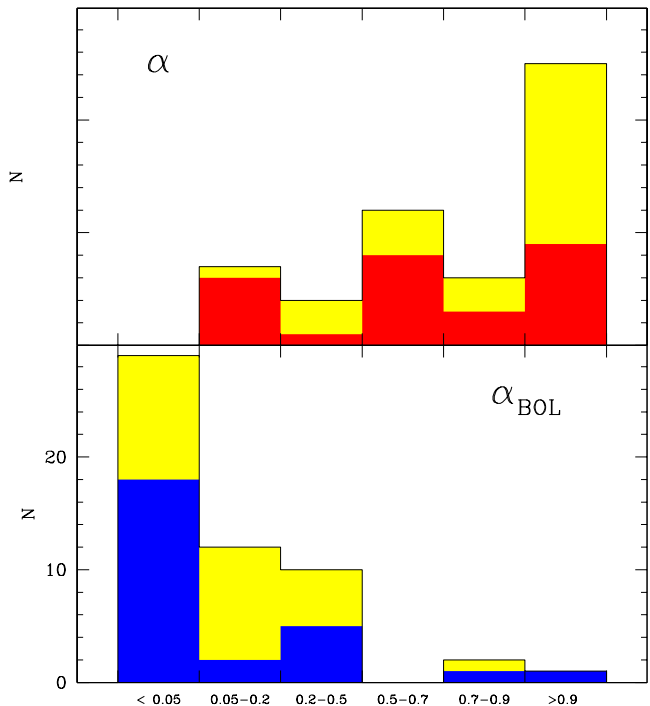}
\caption{ Distribution of the relative AGN contribution to the L-band emission (upper panel)
and to the bolometric emission (lower panel) in our sample of ULIRGs. Blue and red (dark) histograms
refer to LINERs only, while yellow (light) histograms refer to the whole sample.
}
\end{figure}
Aliphatic hydrocarbon absorption features at rest-frame wavelengths 3.4-3.5~$\mu$m have been detected in
seven sources (three AGNs and four LINERs). All these objects have a clear
detection of an AGN according to our model. Moreover, they
are all associated to steep continua, with estimated continuum extinction $\tau>1$ 
(see also Sani et al.~2008).
Therefore, this correlation is a further independent confirmation of the validity of our diagnostics.\\
{\bf Previous L-band diagnostics:} Most of the sources in our sample were presented and
discussed in I06. In this work, the possible presence of the AGN in this
work is estimated through the EW of the 3.3~$\mu$m PAH and the presence of absorption
features. The final classification is shown in
Table~1, and is in general agreement with our results. We stress again
the two main improvements in the analysis presented here: a {\it quantitative} estimate
of the relative AGN/SB contributions, and the inclusion of the continuum reddening as
a further indicator of an obscured AGN (e.g. the source {\bf C} in Fig.~1).\\
{\bf Mid-IR diagnostics:} Imanishi et al.~(2007) presented complete {\em Spitzer}-IRS 
spectroscopy of most of our sources, and an AGN/SB classification based on several
diagnostic features (PAH emission at 6.2, 7.7 and 11.3~$\mu$m, silicate absorption at $\sim$9-10~$\mu$m).
The results are shown in the last column of Table~1, and again are in good agreement 
with the conclusions presented here, with the same limitations discussed above for
previous L-band diagnostics. 

Nardini et al.~(2008) performed an analysis of the same {\em Spitzer} spectra,
focusing on the  5-8~$\mu$m spectral range, and applying an analysis similar to the
one presented here.
The results on the single sources
are in excellent agreement in the two works: this can be verified comparing
our estimates with their Table~1.
Here we only note that the L-band analysis,
besides providing an independent AGN/starburst deconvolution, has a higher diagnostic
power, especially for ULIRGs hosting highly obscured AGNs, thanks to the higher
contrast between AGN and starbursts in the L-band than at 5-8~$\mu$m.

Alternative, powerful mid-IR indicators of the AGN/starburst contributions in ULIRGs
are the silicate absorption features in the 9-10~$\mu$m band (Spoon et al.~2007),
and the presence of high-ionization emission lines (Farrah et al.~2007).
These studies, performed on samples with large overlaps with ours, provide
results in good agreement with our classification. However, as for the other
approaches described in this Section, no quantitative estimates have been 
done based on these indicators.\\
{\bf X-rays:} A strong X-ray emission is one of the strongest AGN indicators.
If the obscuration of the AGN is not too high (column densities $N_H$$<$10$^{24}$~cm$^{-2}$),
the primary AGN spectrum is directly visible at energies below 10~keV, providing
a good estimate of its total luminosity, through bolometric corrections obtained
from AGN SEDs (e.g. Risaliti \& Elvis~2004). In the obscuration is higher, only the
reflection component is visible in the X-rays. In these cases, 
the spectral features of the reflection spectrum can identify the presence of the AGN, however
a good estimate of its total luminosity is not possible, due to the large uncertainty in the
reflection efficiency. 
Extensive studies of local
ULIRGs (e.g. Franceschini et al.~2003, Ptak et al.~2003) show that
the current X-ray observatories can provide useful spectra for 
this diagnostics only for  the 10-20 brighest sources, while most of the
ULIRGs in our sample are too faint for this kind of analysis.
The agreement between X-ray and L-band diagnostics for the few ULIRGs
with high quality X-ray spectra is quite good (Sani et al.~2008, Braito et al.~2009).
However, this analysis cannot be extended to more sources until more sensitive X-ray
observatories are available.

\subsection{Model uncertainties}
The model presented here is based on a series of strong assumptions, which may 
introduce systematic errors, in addition to the statistical errors obtained 
from the fitting procedure. The main possible simplifications, already discussed
in R06, are the assumption of fixed AGN and starburst templates, and the
absence of any absorption/reddening in the starburst component. Other possible relevant 
effects may be caused by different dust extinction curves from the one assumed here.

Regarding possible obscuration of the starburst component, we note that, though
it cannot be excluded, it is not strongly required by the data. Physically, this may
be due to the origin of the starburst emission: the integrated contribution of 
a large number ($>10^6$) of single sources. Each single star can be heavily absorbed 
(as observed, for example, in the central regions of our own Galaxy), but the
single differences are smeared out by the large number of sources, producing a rather
constant spectral profile of the total emission. Regardless of the physical interpretation,
this is also suggested by the observations: the highest quality spectra of pure starburst
are remarkably similar, both in the L-band (R06) and in the 5-8~$\mu$m range (Nardini et al.~2008,
Brandl et al.~2007).
\begin{figure}
\includegraphics[width=8.5cm]{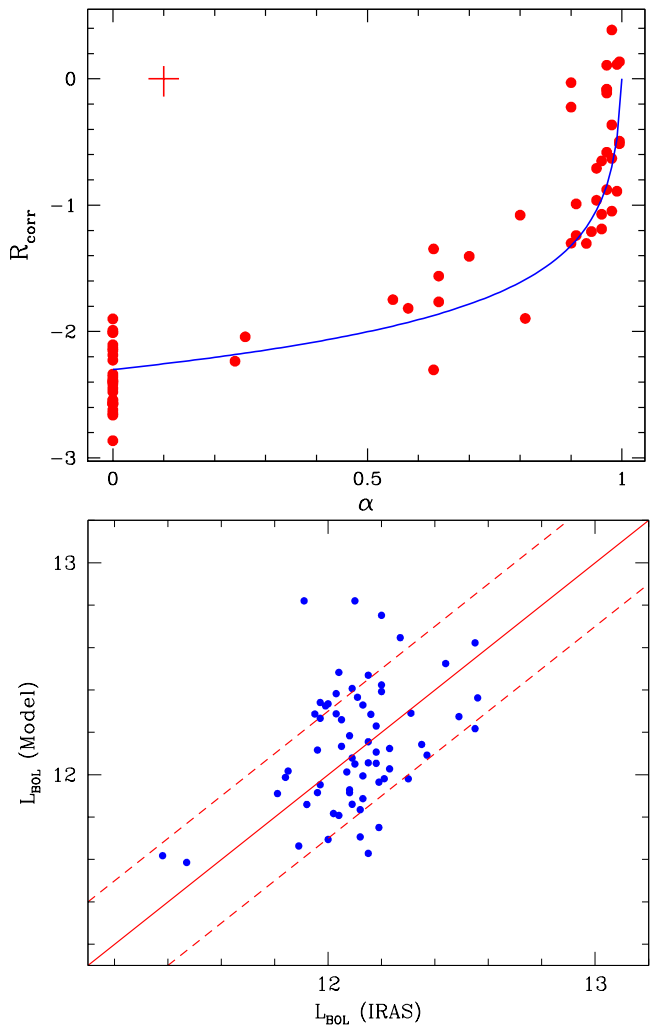}
\caption{ {\em Upper panel:} Ratio between intrinsic L-band and bolometric luminosity 
versus the AGN fraction $\alpha$ obtained from our analysis. The blue continuous line
is that expected based on our model. The cross on the top left corner indicates the average
statistical errors on the two plotted quantities. {\em Bottom panel:} Comparison between the 
estimated and measured bolometric luminosities. The 0.3~dex dispersion around the 1:1 relation
is a fair estimate of the total uncertainty in our prediction of the AGN and SB bolometric
contributions.
\label{isto}
}
\end{figure} 

Regarding the fixed AGN and SB templates, this is an obvious over-simplification, as
shown by the scatter in the spectral parameters (Fig.~2, and R06). We may try to estimate
the effect of this scatter by changing the parameters of the pure AGN and pure SB
templates, and reporting how this affects the estimate of the parameters $\alpha$ and $\tau$.
Since this scatter is hard to measure in a self-consistent way, we prefer to adopt 
a different, observation-based approach. In order to do this, we 
notice that our model can easily provide an estimate of the expected bolometric luminosity
for each source, as a function of $R_{AGN}$, $R_{SB}$, the model parameters $\alpha$ and $\tau$, and
the observed L-band luminosity, $L_{3-4}^{OBS}$. In particular, 
\begin{equation}
L_{BOL}=[\alpha R_{AGN}+(1-\alpha)R_{SB}]\times L_{3-4}^{INTR} 
\end{equation}
where $L_{3-4}^{INTR}$ is
the intrinsic L-band luminosity (i.e. corrected for the extinction of the AGN component) and
is related to the observed one by:
$L_{3-4}^{INTR}$=$L_{3-4}^{OBS}/[\alpha+(1-\alpha)e^{-\tau}]$.
Our estimate of $L_{BOL}$ can then be compared with the observed $L_{BOL}$, as measured by IRAS, by using the
standard calibration of Sanders \& Mirabel~(1996). 
In Fig.~5 we plot two relations:\\
- In the upper panel the ratio $R_{CORR}$ between the absorption-corrected L-band luminosity, $L_{3-4}^{INTR}$,
and the measured bolometric luminosity is plotted versus the L-band AGN fraction, $\alpha$.
The continuous line is the expected relation based on our model, with no free parameters.
In particular, the shape of the curve is determined by the mathematical relations between the
model parameters, while the values at $\alpha$=0 and $\alpha$=1 are the parameters $R_{SB}$ and $R_{AGN}$,
respectively. 
Overall, the agreement between the expected curve and the data is quite good.
This is a strong confirmation of the validity of our method: our simple, two-parameter
model can successfully predict the intrinsic L-band to bolometric ratio.\\
- In the lower panel we directly compare the predicted and measured values of $L_{BOL}$. This
is the most direct way to estimate the overall uncertainties in our analysis. These include both
the statistical errors from the spectral fits, and the systematic errors due to the fixed templates.
The dispersion from the 1:1 relation is lower than 0.3~dex. This is enough to provide
a rough estimate of the AGN contribution in single sources, and an accurate average estimate for the whole sample. 

\section{Conclusions}

L-band spectral analysis of ULIRGs is a powerful way to disentangle
the AGN and starburst contribution to the bolometric emission. This is 
primarily due to the diagnostic power of (a) the 3.3~$\mu$m PAH emission
in starbursts, (b) the continuum slope, indicative of the reddening of the AGN
component, (c) the high ratio (about 100) between the AGN and starburst emission
at 3~$\mu$m, for the same bolometric luminosity.

We applied a simple model to the available L-band spectra of 59 ULIRGs from
the 1-Jy sample, at redshift z$<0.15$, in order to detect and quantitatively estimate
the AGN contribution to the emission of these sources. We find that AGNs are
present in 60\% of ULIRGs, but in most cases they are not the main
energy source. As a consequence, their overall contribution to the total ULIRG
luminosity is $\sim$20\%. We find that the subsample of objects optically classified
as LINERs is rather composite, with similar AGN/SB fractions as the whole ULIRGs
sample.

Our analysis revealed several objects optically classified as starburst or LINERs,
and/or with weak or absent previous evidence of the presence of an AGN in the near/mid-IR,
which according to our model host a powerful, obscured AGN. We plan to observe
these objects in the hard X-rays, in order to obtain an independent, unambiguous confirmation
of these findings.

\section*{acknowledgements}
We are grateful to the referee for his/her constructive comments which significantly helped to improve this paper. 
This work has been partially supported by contract ASI-INAF I/023/05/0.


\end{document}